\documentclass[letter]{aa}
\usepackage{graphicx}
\usepackage{latexsym}
\usepackage{amssymb}
\usepackage{natbib}
\usepackage{txfonts}
\def\teff{\ifmmode T_{\rm eff} \else $T_{\mathrm{eff}}$\fi}

\def\ltsima{$\buildrel<\over\sim$}
\def\lsim{\lower.5ex\hbox{\ltsima}}

\newcommand{\ha}{\ifmmode {\rm H}\alpha \else H$\alpha$\fi}
\newcommand{\hb}{\ifmmode {\rm H}\beta \else H$\beta$\fi}
\newcommand{\lya}{\ifmmode {\rm Ly}\alpha \else Ly$\alpha$\fi}

\newcommand{\ebv}{\ifmmode E_{\rm B-V} \else $E_{\rm B-V}$ \fi}
\def\micron{$\mu$m}
\def\kms{km s$^{-1}$}

\def\ergscm{erg s$^{-1}$ cm$^{-2}$}
\def\msun{\ifmmode M_{\odot} \else M$_{\odot}$\fi}
\def\msunyr{\ifmmode M_{\odot} {\rm yr}^{-1} \else M$_{\odot}$ yr$^{-1}$\fi}
\def\zsun{\ifmmode Z_{\odot} \else Z$_{\odot}$\fi}
\def\lsun{\ifmmode L_{\odot} \else L$_{\odot}$\fi}

\def\mup{\ifmmode M_{\rm up} \else M$_{\rm up}$\fi}
\def\mlow{\ifmmode M_{\rm low} \else M$_{\rm low}$\fi}

\def\aap{A\&A}

\def\apjl{ApJL}
\def\apjs{ApJS}

\newcommand{\oh}{\ifmmode 12 + \log({\rm O/H}) \else$12 + \log({\rm
O/H})$\fi}
\def\Nii{[N\small II]\normalsize $\lambda\lambda$6548,6584}

\def\Siii{[S~{\sc iii}] $\lambda\lambda$9069,9532}

\def\Oiii{[O~{\sc iii}] $\lambda\lambda$4959,5007}

\def\hyperz{{\em Hyperz}}
\def\flyf{\ifmmode f_{\rm Lyf} \else $f_{\rm Lyf}$\fi}
\def\pz{\ifmmode P(z) \else $P(z)$\fi}
\def\ki2{\ifmmode \chi^2 \else $\chi^2$\fi}
\def\zphot{\ifmmode z_{\rm phot} \else $z_{\rm phot}$\fi}

\newcommand{\xphot}{\ifmmode x_\gamma \else $v_\gamma$\fi}
\newcommand{\xobs}{\ifmmode x_{\rm obs} \else $x_{\rm obs}$\fi}
\newcommand{\xcmf}{\ifmmode x_{\rm CMF} \else $x_{\rm CMF}$\fi}
\newcommand{\vexp}{\ifmmode V_{\rm exp} \else $V_{\rm exp}$\fi}
\newcommand{\vmax}{\ifmmode V_{\rm max} \else $V_{\rm max}$\fi}
\newcommand{\nh}{\ifmmode N_{\rm HI} \else $N_{\rm HI}$\fi}
\newcommand{\dv}{\ifmmode \Delta v({\rm em-abs}) \else $\Delta v({\rm em}-{\rm abs})$\fi}

\begin{document}
\title{The impact of nebular emission on the ages of $z \approx 6$ galaxies}

\author{Daniel Schaerer\inst{1,2}
\and
Stephane de Barros\inst{1}
}
\offprints{daniel.schaerer@unige.ch}
\titlerunning{}

\institute{
Geneva Observatory, University of Geneva,
51, Ch. des Maillettes, CH-1290 Versoix, Switzerland
\and
Laboratoire d'Astrophysique de Toulouse-Tarbes, 
Universit\'e de Toulouse, CNRS,
14 Avenue E. Belin,
F-31400 Toulouse, France
}
\date{Received 4 february 2009; accepted 27 april 2009}

\abstract{}{
We examine the influence of nebular continuous and line emission in high redshift star forming galaxies
on determinations of their age, formation redshift and other properties from SED fits. 
}    
{
We include nebular emission consistently with the stellar emission in our SED fitting tool 
and analyse differentially a sample of 10 $z \approx 6$ galaxies in the GOODS-S field studied 
earlier by Eyles et al.\ (2007).
} 
{
We find that the apparent Balmer/4000 \AA\ breaks observed in a number of $z \approx 6$ galaxies
detected at $\ga 3.6$ \micron\ with IRAC/Spitzer can be mimicked by the presence of 
strong restframe optical emission lines, implying in particular younger ages than previously thought.
Applying these models to the small sample of $z \approx 6$ galaxies, we find that this effect
may lead to a typical downward revision of their stellar ages by a factor $\sim 3$.
In consequence their average formation redshift
may drastically be reduced, and these objects may not have contributed to cosmic
reionisation at $z>6$. Extinction and stellar mass estimates may also be 
somewhat modified, but to a lesser extent.}
{Careful SED fits including nebular emission and treating properly
uncertainties and degeneracies are necessary for more accurate
determinations of the physical parameters of high-$z$ galaxies.}


\maketitle
\section{Introduction}
\label{s_intro}
Although star-forming high redshift galaxies have been identified 
out to redshift $z \ga 6$, 
quite little is known about the physical properties of these galaxies,
since this requires in particular sensitive observations in the 
near-IR range and beyond.
With the advent of Spitzer it has recently become possible to detect
galaxies at $z \sim$ 6--7 at wavelengths $\ga$ 3.6 \micron, 
albeit still in quite small numbers
\citep{Egami05,Eyles05,Yan05,Yan06,Eyles07}.
%
One of main surprises of these studies has been the finding of a substantial
Balmer/4000 \AA\ spectral break in a large fraction of these objects,
taken as an indication for the presence of an ``old'' (several hundred Myr)
underlying stellar population, which would indicate very high formation 
redshifts for these galaxies. 
For example, at $z \approx 6$, \citet{Yan06} estimate representative ages of 40--500 Myr
for their sample of $\sim$ 50 IRAC detected objects; \citet{Eyles07}
analyse 10 IRAC detected galaxies and derive ages of $\sim$ 200-700 Myr and 
corresponding $7 \le z_{\rm form} \le 18$.
However, it is possible that the Balmer break is affected or mimicked by emission lines
in high-$z$ galaxies with intense star formation, since their contribution to 
photometric filters increases with $(1+z)$ \citep[see e.g.][]{SP05}.
Determining properties such as maximum ages, past star formation histories,
and formation redshifts is important for a variety of topics including our 
knowledge of galaxy formation, our understanding of cosmic reionisation
and its sources, and others.
It is therefore essential to examine thoroughly the uncertainties in the 
underlying SED analysis.

While possible uncertainties in stellar mass estimates of high-$z$ galaxies
have the focus of several recent discussions \citep[see e.g.][]{Maraston06,Bruzual07,Elsner08},
little attention has been paid to the impact of nebular emission (both
lines and continuous emission) on SED analysis, since these processes
are usually not included in evolutionary synthesis models.
The effect of nebular emission on broad-band photometry of galaxies
has been studied in several papers \citep[e.g.][]{Anders03,Zackrisson08},
and it leads to improvements for photometric redshifts, as shown 
recently by \citet{Ilbert09} and  \citet{Kotulla09met}.
Nevertheless, such models have so far not been applied to analyse the 
properties of high-$z$ galaxies \citep[but see][]{Raiter09}.
In this Letter we examine in a differential manner the effect of nebular
emission on $z \approx 6$ galaxies, and we show that 
it can have a significant impact on determinations of their age and hence 
formation redshift.

In Sect.\ \ref{s_obs} we summarise the galaxy sample and the SED
fitting method. Our results and their implications are presented in
Sect.\ \ref{s_res}. Our main conclusions are discussed and summarised in
Sect.\ \ref{s_conc}. We assume a flat $\Lambda$CDM cosmology with $H_0=70$ \kms\
Mpc$^{-1}$, $\Omega_M=0.3$, and $\Omega_{\rm vac}=0.7$.
\section{Observational data and modeling tools}
\label{s_obs}

\subsection{Selection of $z \approx 6$ galaxies}
To examine the robustness of stellar ages and other physical parameters
of the most distant galaxies, we chose the sample of ten $z \approx 6$ star-forming 
galaxies from the GOODS-South field modeled earlier by \citet[][ hereafter E07]{Eyles07}.
Four of these objects have spectroscopic redshifts determined  from their
\lya\ emission; for the remaining
objects we use the photometric redshifts from the GOODS-MUSIC catalogue adopted by E07.
We use the $i^\prime z^\prime J K_S$, 3.6 \micron, and 4.5 \micron\
ACS, ISAAC, and IRAC photometry from E07.

Since longer wavelength data is also available, we have also used
  for comparison the IRAC photometry at 5.8 and 8.0 \micron\ from the
  GOODS-MUSIC catalogue of \citet{Grazian06} for the subsample of 3
  objects with spectroscopic redshifts also covered by their data.

\subsection{SED fitting tool}
To analyse the SEDs we use a modified version of the \hyperz\
photometric redshift code of \citet{hyperz} described
e.g.\ in \citet{SP05,Schaerer07ero} and adapted 
to include nebular emission.
Among the large diversity of spectral templates included 
in this version, we here use the 2003 GALAXEV synthesis models
from \citet{BC03}, covering different metallicities and a
wide range of star formation (SF) histories (bursts, exponentially
decreasing, or constant SF). 
For comparison with E07, we define the stellar age $t_\star$ as the age since
the onset of star formation, we adopt a Salpeter IMF from 0.1 to 100 \msun,
and we properly treat the returned ISM mass from stars.

To account for the effects of nebular emission from 
young, massive stars on the SED we include nebular emission
(both lines and continua) in a simple manner.
Continuum emission is added to the stellar SED,
as in our synthesis models \citep[cf.][]{SV98,SB99,Scha03}.
The main emission lines of He, C, N, O, S, and other lines
are included using the empirical 
relative line intensities compiled by \cite{Anders03} from galaxies
grouped in three metallicity intervals covering $\sim$ 1/50 \zsun\ to
\zsun.\footnote{See \citet{Kotulla09} for a comparison of the resulting
emission line spectrum with nearby galaxies.}
In addition we include H recombination lines from the
Balmer, Paschen, and Brackett series, as well as \lya;
their relative intensities are taken from \citet{Storey95}
for a typical ISM (density,temperature) of ($n_e=100$ cm$^{-3}$, $10^4$ K).
\footnote{The emissivities of these lines are known to depend little
on $n_e$ and $T$.}
The absolute strength of both the continuous and line emission depend
to first order only on the total number of Lyman continuum photons,
which can be computed from the template stellar SED.
In this manner we include the main nebular emission features from the
UV (\lya) to 2 \micron\ (restframe), necessary to fits the SED of
galaxies at $z>4$ up to 10 \micron\ (IRAC Channel 4).

The free parameters of our SED fits are:
the metallicity $Z$ (of stars and gas),
the SF history described by the timescale
$\tau$  (where the SF rate is SFR $\propto \exp^{-t/\tau}$),
the age $t$, 
the extinction $A_V$ described here by the Calzetti law \citep{Calzetti00},
and whether or not nebular emission is included.
Here we consider three metallicities $Z/\zsun=$1, 1/5, 1/20, a wide
range of $\tau$ values as well as bursts and SFR=const, ages up to the
Hubble time, and $A_V=$ 0--2 mag.
More details on the SED fitting method will be presented in
\citet{Debarros09}.

\section{Results}
\label{s_res}

\begin{figure}[tb]
\includegraphics[width=8.8cm]{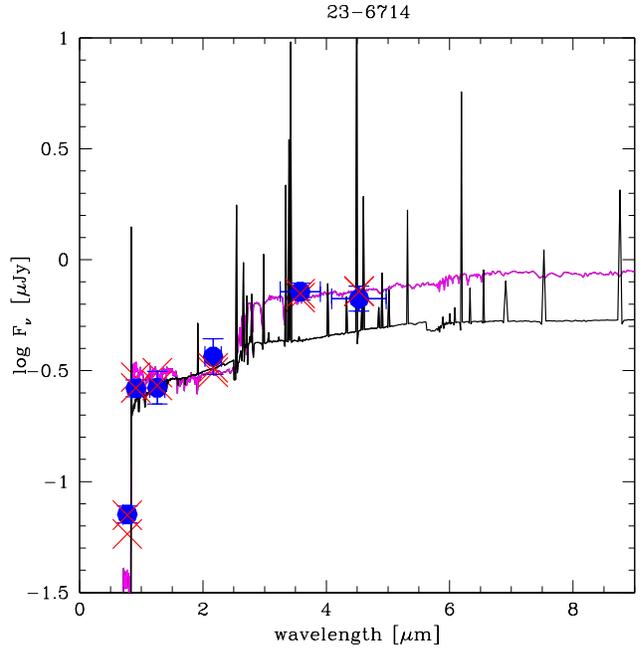}
\caption{
Observed (blue points) and best fit SEDs (solid lines) of the $z=5.83$ galaxy 23\_ 6714 from E07.
The errorbars of the observed wavelength indicate the surface of the normalised
filter transmission curve. Upper limits in flux indicate $1 \sigma$ limits.
Red crosses show the synthesised flux in the filters.
The two SED fits shown are based on a standard Bruzual \& Charlot solar metallicity 
model (magenta), and the same modeling including also nebular emission (black).
While the age of the former (with $\protect\ki2 =1.61$ ) is $\protect\ga 700$ Myr, 
the latter gives $t_\star \sim $ 20 Myr ($\protect\ki2 =0.14$).} 
\label{fig_sed1}
\end{figure}

\begin{figure}[tb]
\includegraphics[width=8.8cm]{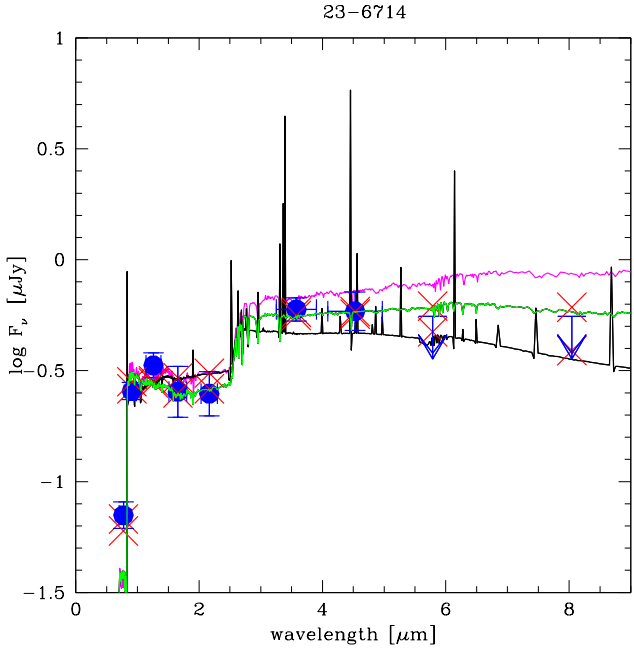}
\caption{Same as Fig.\ 1 for 23\_6714, but including IRAC 5.8 and 8.0
  \micron\ data ($1\sigma$ limits), and using the GOODS-MUSIC photometry.
The magenta line shows the solar metallicity ``standard'' SED fit with an old age
  ($t_\star \protect\ga$ 700 Myr) and no extinction obtained in Fig.\ 1.  
The black line shows our best fit including nebular emission with a very
  young age ($t_\star \sim$ 6 Myr) and $A_V \sim 0.7$, the green line
the $\protect\ga$ 700 Myr best fit with standard (purely stellar) templates.}
\label{fig_sed2}
\end{figure}

\begin{figure}[tb]
\includegraphics[width=8.8cm]{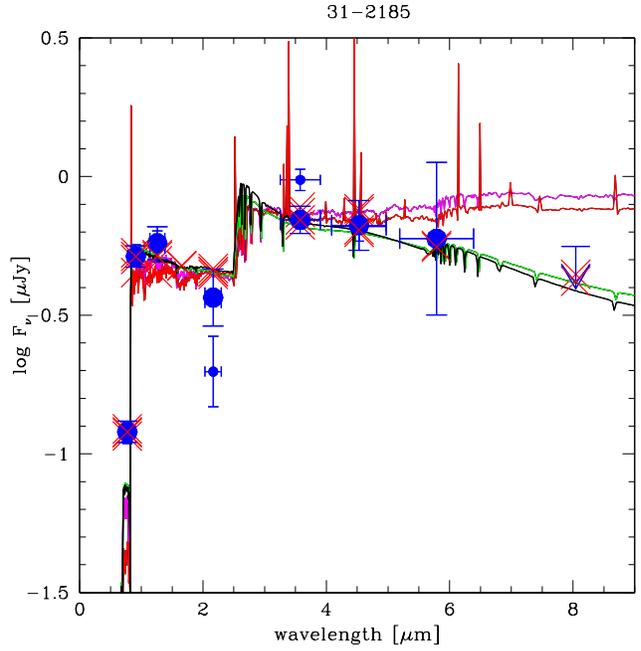}
\caption{Same as Fig.\ 1 for object 31\_2185. The two small blue circles 
illustrated the difference in the photometry of E07 compared to that
of the GOODS-MUSIC catalogue shown with large symbols. 
The red (magenta) line shows the best fit to the photometry of E07 assuming solar 
metallicity models with (without) nebular emission.
The black (green) lines show best fits to the GOODS-MUSIC photometry with (without)
nebular emission.
For both datasets the age obtained with nebular emission is 
$t_\star \sim$ 90--180 Myr, compared to $\sim$ 130--700 Myr with 
standard templates.}
\label{fig_sed3}
\end{figure}

\subsection{Comparison of SED fits with/without nebular emission}
Overall we obtain similar although not identical results to E07, when using the same
assumptions. In other words, adopting solar metallicity and neglecting
nebular emission we find that a significant fraction of the
objects (at least 5/10) are best fit by fairly large ages of the order of $t_\star \ga$
500 Myr and generally very low extinction. The quality of our fits
(expressed in reduced \ki2) is comparable to E07, and
similar values are also found for the stellar masses and current SFR.
For example, the mean age, stellar mass, and extinction of the 10 objects 
we obtain with these assumptions are $\overline{t_\star} \approx$ 500 Myr, 
$\overline{M_\star}=1.2 \times 10^{10}$ \msun, and $\overline{A_V}=0.11$ (i.e.\ $\overline{\ebv} \sim 0.05$).

However, including nebular emission changes quite drastically 
these results. In this case only 1 of 10 objects has a best fit age
$t_\star \ga$ 500 Myr, and the average age of the sample is lowered by a factor
$\sim 4$ to $\overline{t_\star} \approx$ 120 Myr. A slightly lower mean stellar mass
$\overline{M_\star}=7.8 \times 10^{9}$ \msun, and a higher extinction $\overline{A_V}=0.34$ 
$\overline{\ebv} \sim 0.08$) is also obtained. 
Compared to the above set of models
the \ki2\ values is found to be lower for half of the objects. 
A more rigorous comparison, including also a careful discussion of the uncertainties,
 is deferred to a later paper.

Examples of SED fits for two objects with known \lya\ emission and known
spectroscopic redshift are show in Figs.\ 1 to 3.
Figure \ref{fig_sed1} shows the best fit solution
using pure stellar SEDs at solar metallicity compared to the best fit
including nebular emission (found to be for $Z=1/5$ \zsun). While the
age of the latter is $t_\star \approx 20$ Myr, the former has an age
close to the maximum allowed for the redshift of this object
($z=5.8$). With nebular emission the apparent Balmer break is mimicked
by the presence of strong restframe optical emission lines boosting
the flux both in the 3.6 and the 4.5 \micron\ filters\footnote{The
  strongest lines in the 3.6 \micron\ filter are \Oiii\ and H$\beta$,
  the strongest at $\sim 4.5$ \micron\ are H$\alpha$, \Nii, and
  \Siii.}. 
For this object (23-6714) a non-zero extinction, of the order of
$A_V \sim$ 0.2--0.8 depending on metallicity, is required.

Since nebular lines at this redshift are expected to affect less the
photometry at longer wavelenghts ($\ga$ 5 \micron), it is interesting
to examine SED fits including also the available 5.8 and 8.0 \micron\ data
(see  Fig.\ \ref{fig_sed2}). For consistency we therefore use the 
GOODS-MUSIC photometry in all bands. A good agreement with the observations
is found, and again significantly lower ages are obtained with nebular 
emission. For 23-6714 we find the best fit with $Z=1/5$ \zsun, a very young 
age of $t_\star \sim 6$ Myr and $A_V \sim 0.7$, compared to ages of 
$\ga$ 700 Myr with standard (purely stellar) templates.

Figure \ref{fig_sed3} shows an object (31-2185 from E07) with an apparent 
Balmer break of similar strength. An old age $t_\star \sim$ 700 Myr is again
obtained here for standard stellar SED fits, whereas the inclusion of nebular
emission lowers the age to 90--180 Myr, depending if we include or not the
photometry in Channels 3 and 4. For this object and for the GOODS-MUSIC
photometry it turns out that nebular emission is not crucial, since relatively
short star-formation time scales and hence solutions with few ionising stars
are preferred.

The above results clearly illustrate that it is important to 
include the effects of nebular emission (lines in particular) to
determine the properties of these high-$z$ galaxies, most importantly
their stellar ages.

\subsubsection{Comparison with earlier work}
Why does this conclusion disagree with E07 who also examined the
possibility of ``line contamination'' in the IRAC filters, which are
crucial to measure the Balmer break or its absence?  There are three
main reasons for it.
First, E07 compute the contribution of \ha\ to the 4.5
\micron\ filter from an estimate of the unobscured SFR(UV), which
assumes a long ($\ga 100$ Myr) SF timescale. However, for shorter
timescales the \ha\ luminosity can be larger by up to a factor $\sim$
4, as illustrated e.g.\ by \citep[][see their Fig.\ 15]{Verh08}.
Second, as noted by E07 the contribution of \ha\ is further enhanced 
when extinction is taken into account.
Finally, \Oiii\ plus \hb\ have strong enough intrinsic intensities
(typically 1.8--2.2 times \ha) to boost as well the flux in the 3.6 \micron\ 
filter, thereby mimicking a ``Balmer break'' between IRAC Channels 1-2
and shorter wavelength filters.
When the effects of nebular emission lines are taken into account in a
consistent manner with the ionising flux predicted by the SF history
and when extinction is allowed for, we find that -- at $z \sim 6$
considered here -- their contribution to IRAC channels 1 and 2 can be
significant\footnote{See also the predictions of \citet{Zackrisson08}
  who also find significant line contamination even for constant SF
  over 50 Myr.}.  This largely explains our lower age estimates
compared to the earlier work of E07.

\subsubsection{How to distinguish ``old'' and young populations?}
Can the two type of SED fits, explaining the Balmer break
with ``old'' stars or by the presence of emission lines, be
distinguished observationally?

For the SED fit of 23-6714 shown in Fig.\ \ref{fig_sed1} for example,
the current SFR $\sim 800$ \msunyr. From this the expected \ha\ flux
is of the order $3.\times 10^{-16}$ \ergscm, clearly beyond the reach
of current spectroscopic facilities at$\sim$ 4.5 \micron. 
This object being the brightest of the sample it appears that 
direct detections of the restframe optical emission lines
need to await future facilities, such as the JWST.

In principle deep photometry at longer wavelengths, where the 
expected strength of emission lines decreases, could help to disentangle
the two solutions. However, the current 5.8 and 8.0 \micron\ data 
from Spitzer is not deep enough to rule out the ``old'' SEDs, as
shown for the two brightest 
objects 23-6714 and 31-2185. Although the spectrum of the $\sim  700$ Myr old 
population displays an excess with respect to the 5.8 and/or 8.0 \micron\ fluxes,
its significance is $\la 2 \sigma$ at best. 
Future, deeper observations should be able to set firmer limits
on this issue.

In any case we note that neglecting nebular emission for high-$z$ 
galaxies with ongoing massive star formation --- as testified e.g.\ by their
\lya\ emission often used to confirm their redshift spectroscopically --- is physically 
inconsistent and its effect on the determination of their 
physical parameters should be taken into account.

\subsection{Implications}
As shown above, the inclusion of nebular emission (lines and continuum)
can alter the physical properties of galaxies determined from
their SED.
For example, considering just the average properties of our best
fit solutions for the 10 objects fitted here we obtain 
$\overline{t_\star} \approx$ 400 Myr, $\overline{M_\star}=1.1 \times 10^{10}$ \msun, and
$\overline{A_V}=0.16$ for standard Bruzual \& Charlot spectral templates when
metallicity is also varied.
Including nebular emission we obtain $\overline{t_\star} \approx$ 120 Myr,
$\overline{M_\star}=7.9 \times 10^{9}$ \msun, and a higher extinction $\overline{A_V}=0.34$.
Although indicative of the trend obtained with nebular emission, the reader
should be aware that large deviations are obtained and uncertainties
should be treated properly (cf.\ Sect.\ \ref{s_discuss}).

Clearly the most striking result is that the average age of the $z\approx 6$
galaxies may be decreased typically by a factor $\sim 3$ compared to earlier studies
\cite[cf.][]{Yan06,Eyles07}.
Translated to the average redshift of formation
$z_{\rm form}$ this would imply a shift from $z_{\rm form} \sim$ 9. to 6.6, 
assuming $\overline{z}=5.873$ for the average redshift of our sample (cf.\ E07).
For only one out of 10 objects do we find best fit ages of $>200$ Myr, 
i.e.\ a formation redshift beyond 7.
If true, this means in particular that the contribution of the galaxies currently
observed at $z \sim 6$ to cosmic reionisation must have been
negligible at $z \ga 7$, in contrast to the finding of E07.

The average extinction of $A_V \sim 0.34$ mag (\ebv $\sim$ 0.08) we
find, corresponding to a UV attenuation by a factor $\sim 2.2$ is also
worth noticing.  With standard spectral templates, neglecting nebular
emission, the average is $A_V=0.16$. This result indicates that dust
attenuation may be larger than previously thought for Lyman Break galaxies at this
redshift. For comparison, \citet{Bouwens08} assume a UV attenuation
factor of $\sim$ 1.5 at $z \sim 6$.
If true and representative for the population of $z \sim 6$ galaxies,
it may imply an upward revision of the SFR density by $\sim 50$ \%.
Finally stellar masses and hence the estimated stellar mass density
of i-drop galaxies may also be slightly affected. Also already mentioned,
the average stellar mass of our sample $\overline{M_\star}$ is $\sim 30$ \% lower
when we include nebular emission in the SED fits and leave the other
assumptions unchanged. 

\section{Discussion and conclusions}
\label{s_discuss}
\label{s_conc}

The main objective of this work has been to examine the
effect of nebular emission (lines and continua) on the derivation
of physical parameters of $z \sim 6$ galaxies through broad band SED fits.
Although our work clearly indicates that this effect can significantly
alter derived properties such as galaxy ages and extinction, we do not 
(yet) aim at determining absolute values.

Indeed both uncertainties/difficulties in the photometry as well
as uncertainties and degeneracies in the models remain and their
effect needs to be quantified properly.
For example, the differences obtained using photometry from two sources 
(revisions) has been illustrated above.
Stellar ages and other quantities may also vary if different,
more complex SF histories, different extinction laws (possibly also
extinction differences between stars and the gas), different IMFs 
are allowed in the models.

Last but not least, our and other SED fitting tools are also limited by 
their dependence on stellar tracks, which are e.g.\ sensitive
to input physics such as the treatment of rotation \citep{Vazquez07}, 
to the treatment of advanced phases such as TP-AGB stars \citep[cf.]{Maraston06}
and others.
Once the input physics of the SED fitting code and the photometric
data is fixed, errors and degeneracies in the fit parameters need
to quantified in detail and 
for each individual object.
Finally appropriate quantitative methods should be used to
determine the average properties and their uncertainties for samples
of galaxies. Such detailed analysis, clearly beyond the scope of
this Letter, will be presented in a forthcoming publication
\citep{Debarros09}.

Independently of these remaining uncertainties, our work shows clearly
that the presence of strong Balmer/4000 \AA\ breaks in $z \approx 6$ galaxies
is not necessarily due to stars, but can also be mimicked by restframe 
optical nebular emission lines, whose contribution to broad band filters
increases with $(1+z)$. Apparently ``old'' ages and high formation 
redshifts of galaxies with strong ongoing star formation may need 
to be revised significantly downward.
Finally, the impact of nebular emission on SED analysis of other galaxies, also
at lower redshift, needs also to be examined systematically.

\acknowledgements
We would like to thank Matthew Hayes for interesting discussions,
Ana Lalovic for the exploration of other SED fits, Roser 
Pell\'o for regular discussions and support with \hyperz,
and David Elbaz for help with GOODS data.
This work was supported by the Swiss National Science Foundation.
\bibliographystyle{aa}
\bibliography{references}

\end{document}